\documentclass[10pt,pra,twocolumn,amssymb,nofootinbib,showpacs]{revtex4}
\usepackage{bm}
\usepackage{graphicx}
\usepackage{amsmath}
\usepackage{bbm}

\begin{document}
\title{\bf Numerical analysis of a spontaneous collapse
model for a two--level system}
\author{Angelo Bassi}
\email{bassi@ictp.trieste.it} \affiliation{The Abdus Salam
International Centre for Theoretical Physics, Trieste, and \\
Istituto Nazionale di Fisica Nucleare, sezione di Trieste, Italy.}
\author{Emiliano Ippoliti}
\email{ippoliti@ts.infn.it} \affiliation{Department of Theoretical
Physics, University of Trieste,  and \\ Istituto Nazionale di
Fisica Nucleare, sezione di Trieste, Italy.}
\begin{abstract}
We study a spontaneous collapse model for a two--level (spin)
system, in which the Hamiltonian and the stochastic terms do not
commute. The numerical solution of the equations of motions allows
to give precise estimates on the regime at which the collapse of
the state vector occurs, the reduction and delocalization times,
and the reduction probabilities; it also allows to quantify the
effect that an Hamiltonian which does not commute with the
reducing terms has on the collapse mechanism. We also give a clear
picture of the transition from the ``microscopic'' regime (when
the noise terms are weak and the Hamiltonian prevents the
state vector to collapse) to the ``macroscopic'' regime (when the
noise terms are dominant and the collapse becomes effective for
very long times). Finally, we clarify the distinction between
decoherence and collapse.
\end{abstract}
 \maketitle

\section{Introduction}

Stochastic Schr\"odinger equations find fruitful applications both
in the theory of open quantum systems \cite{da,lib,pk,bas,peo} and
within collapse models
\cite{gpr,per,pp,di,bel,bar,bla,jp,ip,ad1,ad2,rep}. In the first
case, the stochastic terms mimic the effect of the environment on
the open system\footnote{The advantage of this kind of approach
over the standard one based on the reduced density matrix of the
open system is mainly computational: in some physical situations,
it turns out to be computationally easier to solve numerically the
stochastic Schr\"odinger equation and to perform the appropriate
averages over different trajectories, rather than solving
numerically the equation for the statistical operator.}. In the
second case, non--linear and stochastic terms are added to the
standard Schr\"odinger equation to reproduce --- under appropriate
conditions --- the collapse of the wavefunction\footnote{More
specifically, it has been shown \cite{gpr} that a particular
choice of the stochastic terms exists such that: 1) the quantum
properties of microscopic systems are not altered in any
appreciable way; accordingly, these models reproduce all
experimental results known at present; 2) at the macroscopic
level, the reduction mechanism becomes rapidly effective so that
macro--objects are always localized in space. Note that in the
literature, little attention (see, e.g., \cite{per}) has been
given to the behavior of the stochastic dynamics when the
Hamiltonian and the stochastic terms do not commute; here we will
extensively discuss the consequences of this noncommutativity.};
in this way, it is possible to describe within one single
dynamical equation both the quantum properties of microscopic
systems and the classical properties of macroscopic objects, and
in particular the outcomes of measurement processes.

In this paper, we analyze with numerical techniques the dynamics
described by the following stochastic Schr\"odinger equation in
the Hilbert space ${\mathbb C}^{2}$:
\begin{eqnarray} \label{sm}
d|\psi_{t}\rangle & = & \left[ -i\omega\sigma_{x} dt \, + \,
\sqrt{\gamma}(\sigma_{z} - \langle \sigma_{z} \rangle)\, dW_{t}
\frac{ }{ }\right.
\nonumber \\
& & \left.-\frac{\gamma}{2}(\sigma_{z} - \langle \sigma_{z}
\rangle)^{2}\, dt\, \right] |\psi_{t}\rangle,
\end{eqnarray}
which is nonlinear and, as a consequence, nonunitary; anyway,
it is easy to check that it preserves the norm of
$|\psi_{t}\rangle$. $\langle \sigma_{z} \rangle$ denotes the
quantum average $\langle \psi_{t}| \sigma_{z} |\psi_{t}\rangle$,
while $\sigma_{x}$, $\sigma_{z}$ are the usual spin operators.
$W_{t}$ is a standard Wiener process: ${\mathbb E}\, [dW_{t}] = 0$
and ${\mathbb E}\, [dW_{t}^{2}] = dt$, and the (positive)
parameter $\gamma$ measures the strength of the coupling between
the spin operator $\sigma_{z}$ and $W_{t}$. Finally, the
stochastic dynamics is defined on a probability space $(\Omega,
{\mathcal F}, {\mathbb  P})$ with the filtration $\{ {\mathcal
F}_{t}: t \in [0, \infty) \}$ generated in a standard way by the
Wiener process $W_{t}$.

Eq. (\ref{sm}) describes the evolution of the spin--wavefunction
of a $1/2$ spin particle; the dynamics is non--trivial, due to the
presence of two competing terms which do not commute: on the one
side the Hamiltonian $H = \hbar \omega\sigma_{x}$, whose effect is
to rotate the state vector along the $x$ axis of the Bloch sphere;
on the other side, the spin operator $\sigma_{z}$ which tends to
localize the state vector along one of its two eigenstates
$|+\rangle$, $|-\rangle$.

We will begin our analysis by studying the equation for the
statistical operator $\rho(t)$ --- describing the ensemble of
states each of which evolves according to Eq. (\ref{sm}) --- which
in this case can be solved exactly (section II); we will give a
qualitative description of the evolution of $|\psi_{t}\rangle$,
based on the dynamics for $\rho(t)$. We will next solve Eq.
(\ref{sm}) numerically, showing that a much richer quantity of
information can be obtained: in particular, we will analyze the
regime in which the reduction of the state vector to one of the two
eigenstates of $\sigma_{z}$ occurs, and we will clarify the
distinction between collapse and decoherence (section III); we
will analyze the reduction probabilities (section IV), the
delocalization mechanism (section V), and the transition from the
``quantum'' ($\gamma \ll \omega$) to the ``classical'' ($\gamma
\gg \omega$) regime (section VI). We will conclude our analysis
with a comparison between the model here analyzed and
space--collapse models (section VII).

\section{The statistical operator}

The statistical operator $\rho(t) = {\mathbb E}\,
[|\psi_{t}\rangle\langle \psi_{t}|]$ obeys the equation of
motion\footnote{The dynamical evolution embodied in Eq.
(\ref{eso}) has been studied in detail in Ref. \cite{per}.}:
\begin{equation} \label{eso}
\frac{d}{dt}\,\rho(t) \; = \; -i\omega\, \left[ \sigma_{x},
\rho(t) \right] \, - \, \gamma \left[ \rho(t) - \sigma_{z} \rho(t)
\sigma_{z} \right],
\end{equation}
which of course is of the Lindblad--type \cite{lind}. It can be
solved analytically for any given initial condition; if we
consider the matrix elements of $\rho$ with respect to the
eigenstates $|+\rangle$, $|-\rangle$ of $\sigma_{z}$:
\[
\rho  \; \longrightarrow \; \left(
\begin{array}{cc}
\langle +| \rho |+\rangle & \langle +| \rho |-\rangle \\
\langle -| \rho |+\rangle & \langle -| \rho |-\rangle
\end{array} \right) \; = \; \left(
\begin{array}{cc}
x & y + iz \\
y -iz & 1 - x
\end{array} \right),
\]
we get the following equations for $x$, $y$ and $z$:
\begin{equation} \label{eqxyz}
\left\{
\begin{array}{lll}
\dot{x}(t) & = & -2\omega\, z(t) \\
\dot{y}(t) & = & -2\gamma\, y(t) \\
\dot{z}(t) & = & -\omega + 2\omega\, x(t) -2\gamma\, z(t).
\end{array}
\right.
\end{equation}
The complete solutions of equations (\ref{eqxyz}) are written in
appendix I; here we limit ourselves to discuss some general
features.

The equation for $y(t)$ is, trivially, a decaying exponential; the
two equations for $x(t)$ and $z(t)$ lead to the following second--order 
differential equation for $z(t)$:
\begin{equation}
\ddot{z}(t) \; + \; 2\gamma\, \dot{z}(t) \; + \; 4\omega^{2}\,
z(t) \; = \; 0,
\end{equation}
which is the equation of a damped harmonic oscillator;
accordingly, for any value of $\omega$ and $\gamma$ (different
from zero) the off--diagonal elements of the density matrix
decrease exponentially in time. This results would suggest that
reductions occur for {\it any} value of $\gamma$ --- even though,
of course, we expect the localization mechanism to be less
efficient for smaller values of $\gamma$ --- but we will see in
the following sections that the real situation is subtler.

The time evolution of $x(t)$ has the following interesting
property: for any value of $\omega$ and $\gamma$ (different from
zero) and for any given initial condition, one has
\begin{equation} \label{aslim}
x(t)\xrightarrow[t \,\rightarrow \, +\infty]{}\frac{1}{2}
\end{equation}
This result is interpreted in the following way: suppose that the
initial (at time $t = 0$) state vector is, e.g.:
\begin{equation} \label{ivs}
|\psi_{0}\rangle \; = \; \sqrt{\frac{3}{4}}\, |+\rangle \; + \;
\sqrt{\frac{1}{4}}\, |-\rangle,
\end{equation}
and assume that after a certain amount of time it has been
stochastically reduced either to the state $|+\rangle$ or to the
state $|-\rangle$, with the correct quantum probabilities. As a
consequence, the density matrix becomes almost diagonal, with the
diagonal elements equal to $3/4$ and $1/4$.

Result (\ref{aslim}) states that, as time goes to infinity, the
diagonal elements must change and become asymptotically equal to
$1/2$: this implies that, however it has been initially reduced,
the state vector starts jumping between the two eigenstates, so
that after a certain amount of time there is an equal probability
of finding it either reduced around $|+\rangle$ or around
$|-\rangle$, despite that, at the very beginning, there was a much
higher probability to find it reduced around $|+\rangle$. In the
following sections we will give a quantitative analysis of this
effect.

Note that the above result does not come as a surprise, since a
general theorem by Spohn \cite{sp} assures that, in the 
finite--dimensional case, when an equation of the quantum dynamical
semigroup type admits a steady solution, then any other solution
converges to it for $t \, \rightarrow \, +\infty$. In our case,
equation (\ref{eso}) has $\frac{1}{2}I$ as the only steady
solution, where $I$ is the identity matrix: accordingly, the
off--diagonal elements must necessarily vanish as time increases,
while the diagonal elements must converge to $1/2$.

\section{Reduction times}

The dynamics described by eq. (\ref{sm}) has several interesting
properties. The first quantity which we have analyzed is the
reduction time, i.e. the time it takes for the state vector to
reduce to one of the two eigenstates $|+\rangle$, $|-\rangle$ of
the spin operator $\sigma_{z}$. A {\it reduction process} must
satisfy the following two requirements:
\begin{enumerate}
\item The state vector $|\psi_{t}\rangle$ must get ``close enough''
to one of the eigenstates of the reducing operator, $\sigma_{z}$
in our case;

\item Once the state vector reaches such an eigenstate, it must
remain close to it for a ``sufficiently long'' time interval.
\end{enumerate}
Let us make these two requirements more precise. When
$|\psi_{t}\rangle$ gets close to the eigenstate $|+\rangle$
($|-\rangle$), the square modulus $|\langle + |\psi_{t}\rangle|^2$
(respectively, $|\langle - |\psi_{t}\rangle|^2$) approaches the
value 1: accordingly, the first requirement amounts to saying that
either $|\langle + |\psi_{t}\rangle|^2$ or $|\langle -
|\psi_{t}\rangle|^2$ is greater than $1 - \varepsilon$, where
$\varepsilon$ is a parameter depending on the physical situation
under study. Following the second requirement, when --- let us say
--- it happens that $|\langle + |\psi_{t}\rangle|^2 > 1 -
\varepsilon$, a reduction has occurred only if this condition
holds for a time interval of length $\tau$; once more, the
suitable value for $\tau$ depends on the specific physical system.

According to the above arguments, we define the {\it reduction
time} as follows: it is the smallest time $t_{r}$ such that, for
$t \in [t_{r}, t_{r} + \tau]$, the square modulus $|\langle +
|\psi_{t}\rangle|^2$ (or $|\langle - |\psi_{t}\rangle|^2$) is
greater than $1 - \varepsilon$.

\begin{figure}
\begin{center}
{\includegraphics[scale=0.41]{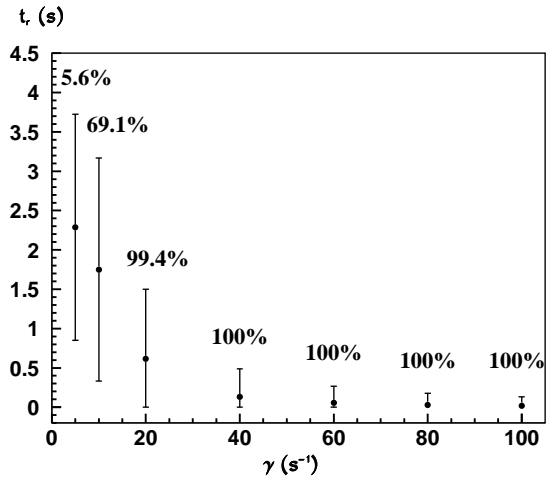}} \caption{Reduction time
$t_{r}$ as a function of $\gamma$; the simulations have been made
for $\gamma = 5,10,20,40,60,80,100$ s$^{-1}$, while $\omega = 1$
s$^{-1}$. The picture shows the average values and the standard
deviations over 100000 trajectories. Above each bar, we have
indicated the fraction of the total trajectories which have been
reduced within the time interval $[0, 2\pi]$ seconds.}
\label{trid}
\end{center}
\end{figure}

Fig. \ref{trid} shows the results of the numerical simulations for
the reduction time as a function of $\gamma$, with the initial
state vector\footnote{In all simulations, we have taken (\ref{ivs})
as the state vector at time $t=0$.} $|\psi_{0}\rangle$ as in eq.
(\ref{ivs}) and $\omega = 1$ s$^{-1}$.
We have taken $\varepsilon = 1/100$, while $\tau$ has been chosen
according to the following criterion: when $\gamma = 0$ and
$|\psi_{0}\rangle = |+\rangle$ or $|-\rangle$, the state vector
rotates between the two eigenstates of $\sigma_{z}$ with frequency
$\omega$. Accordingly, even if no reducing terms are present, the
dynamics brings periodically $|\psi_{t}\rangle$ close to, e.g.,
$|+\rangle$, in such a way that $|\langle + |\psi_{t}\rangle|^2$
remains greater than $1 - \varepsilon$ for a time equal to: $\pi/2
- \arcsin (1 - \varepsilon) \simeq 0.141$ s. We have then chosen
$\tau$ to be equal to $10 \times [\pi/2 - \arcsin (1 -
\varepsilon)]$, to distinguish a reduction from the effects of the
oscillatory motion induced by the Hamiltonian $H$.

The simulation shows that, as expected, the reduction time
decreases for increasing values of $\gamma$. When $\gamma$ is
sufficiently larger than $\omega$ ($\gamma \geq 60$ s$^{-1}$), the
reduction time becomes a well--defined quantity: all trajectories
are reduced, and the variance associated to $t_{r}$ decreases as
$\gamma$ increases\footnote{The mathematical proof that, for
$\gamma \gg \omega$, the state vector is reduced to one of the
eigenstates of the ``reducing operator'' ($\sigma_{z}$, in our
case), was first given in ref. \cite{gpr}. See also ref.
\cite{ad2} for a martingale proof.}. On the other hand, when
$\gamma$ is of the same order of --- or smaller than --- $\omega$,
only a few trajectories (or none) are localized and the variance
associated to the reduction times, increases: in this regime, the
reduction process progressively disappears. This result is
particularly important in relation to the time evolution of the
off--diagonal elements of the density matrix.

In the previous section, we have seen that, for {\it any} value of
$\gamma$, the off--diagonal elements of the density matrix
decrease exponentially in time. According to the numerical
simulations of Fig. \ref{trid}, such a damping has two different
origins according to whether $\gamma$ is significantly greater
than $\omega$ or not: in the first case, the damping is due to the
localization process; in the second case, it is a consequence of a
phase randomization (decoherence) of the different trajectories.

\begin{figure}
\begin{center}
\makebox[4.2cm]{{\includegraphics[scale=0.23]{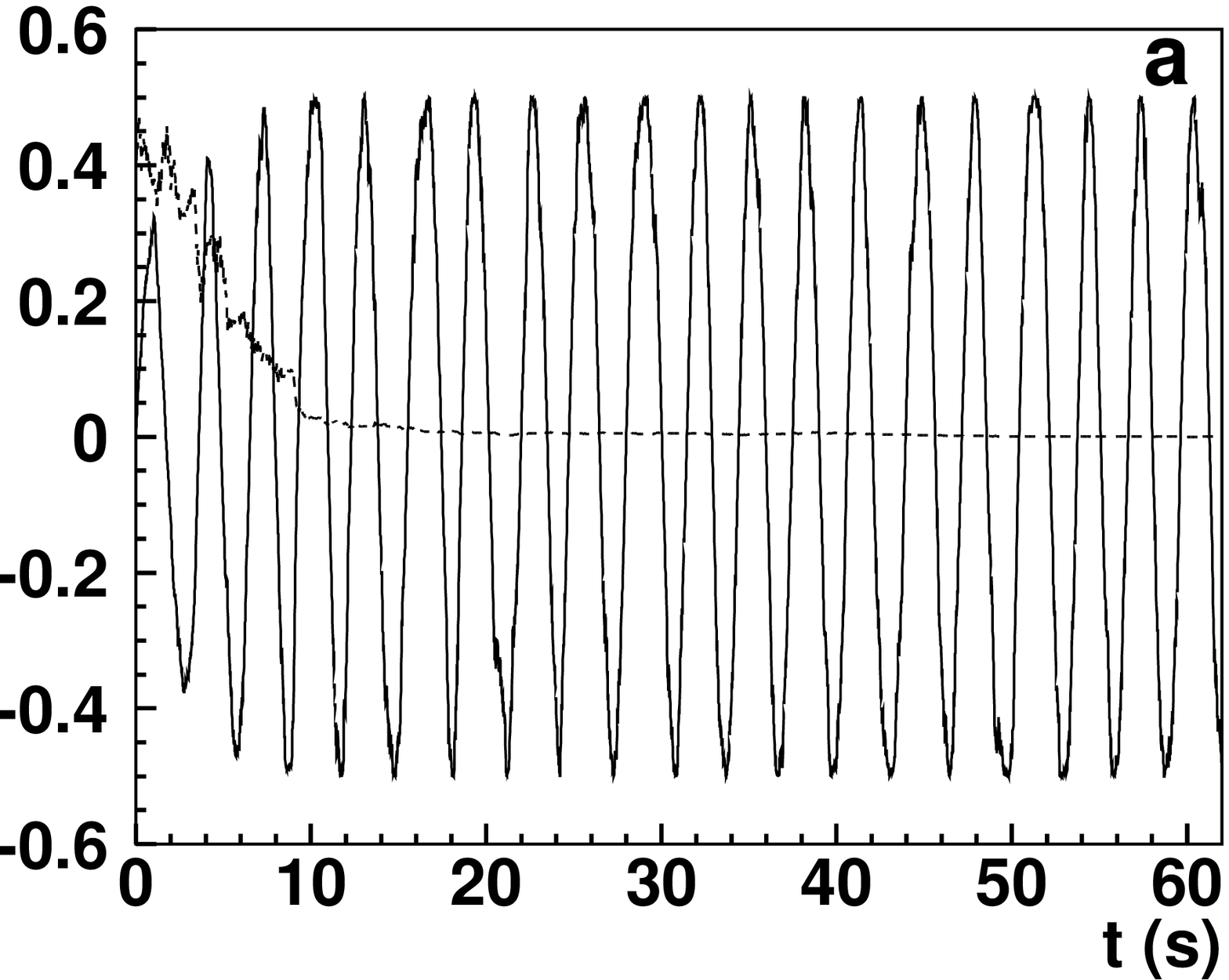}}}
\makebox[4.2cm]{{\includegraphics[scale=0.23]{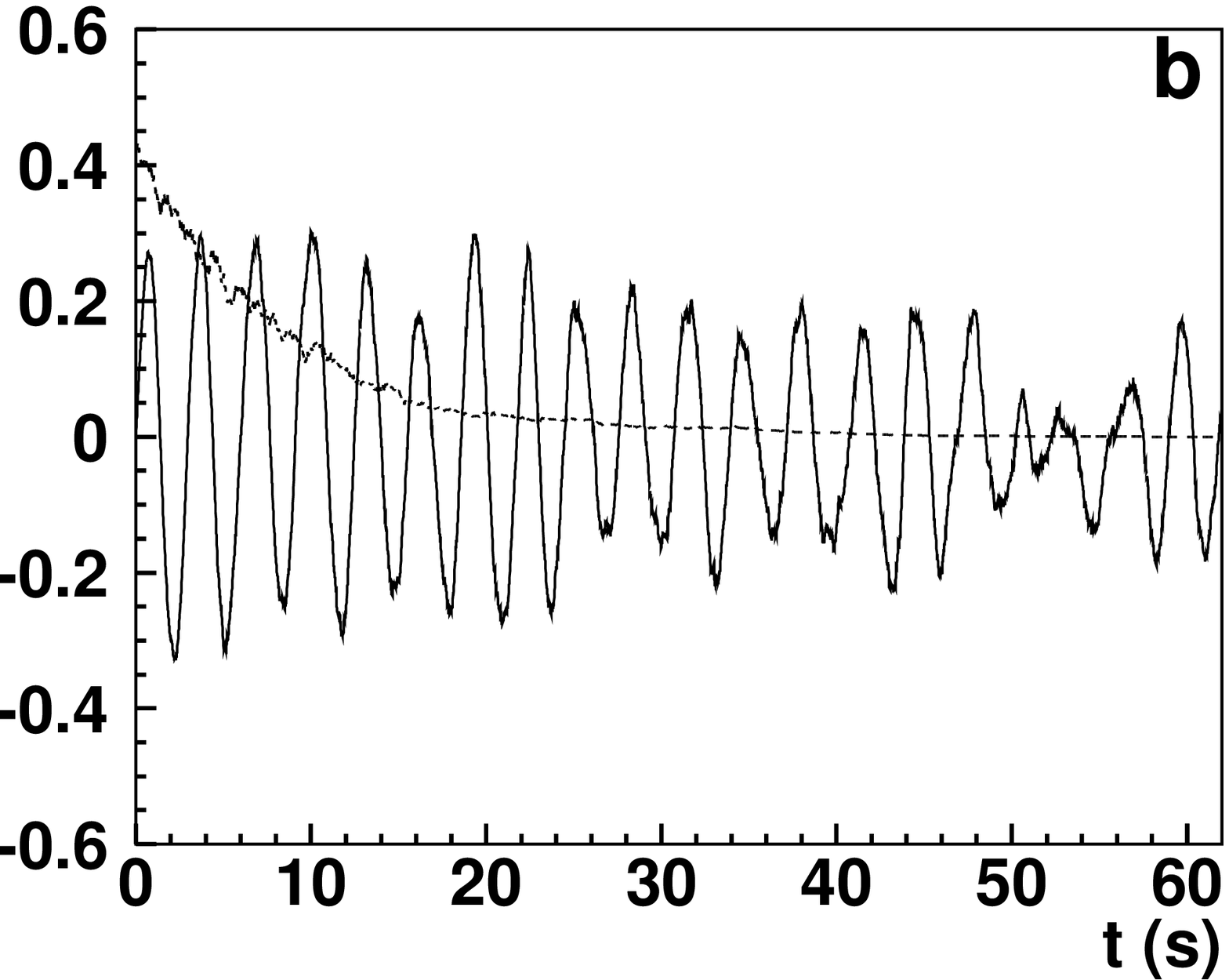}}}
\makebox[4.2cm]{{\includegraphics[scale=0.23]{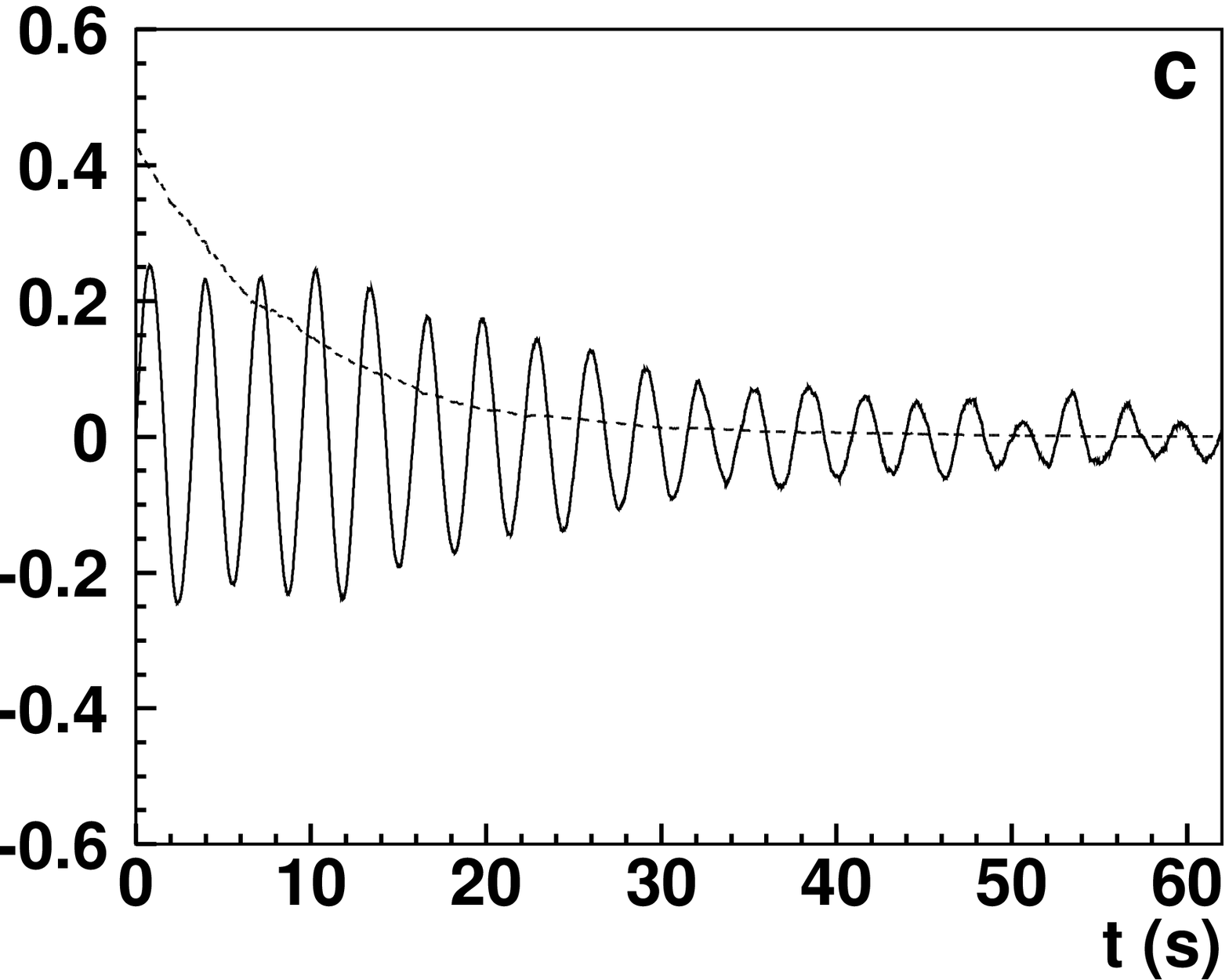}}}
\makebox[4.2cm]{{\includegraphics[scale=0.23]{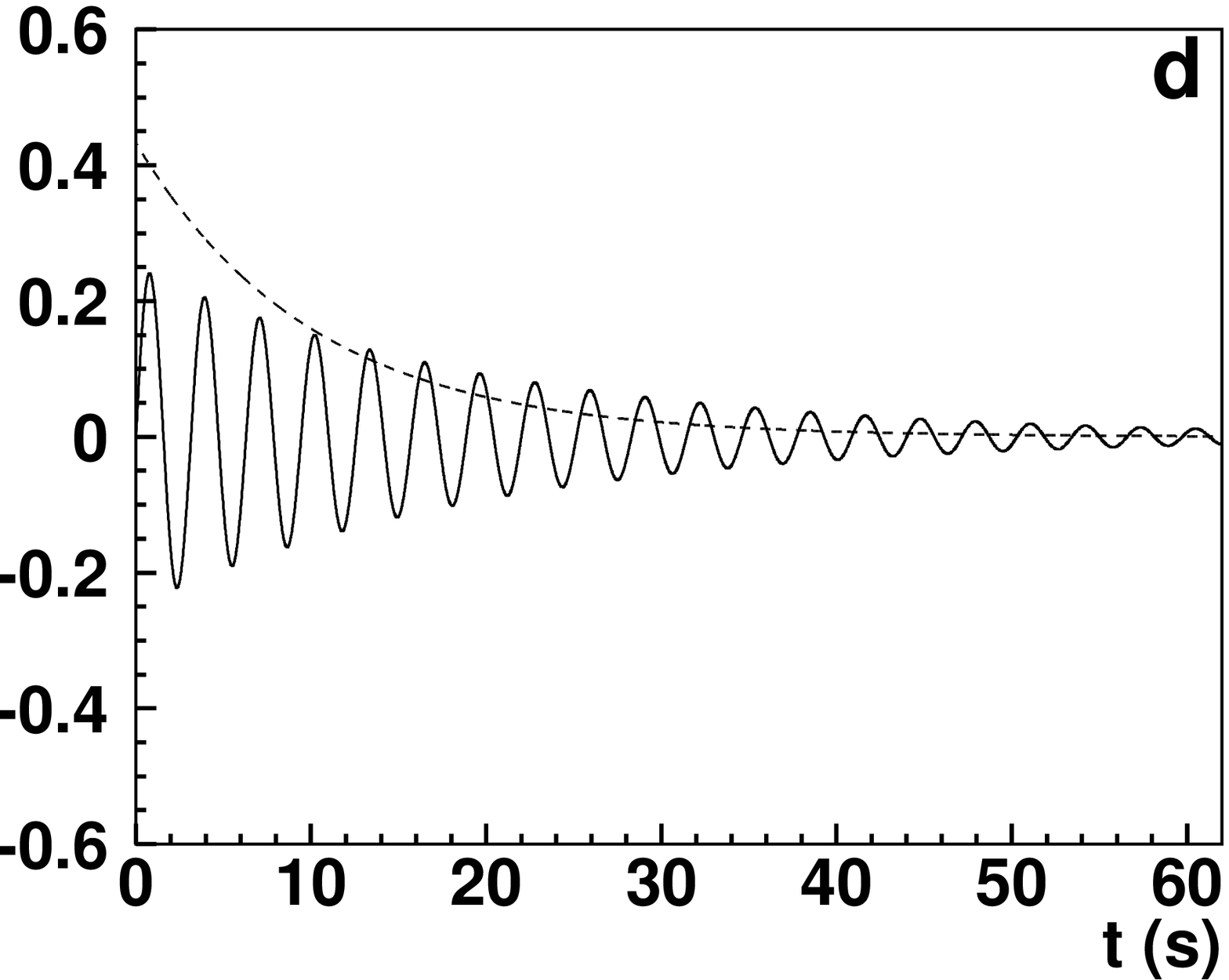}}}
\caption{a) Time evolution of the real (dotted line) and imaginary
(continuous line) parts of $\langle +
|\psi_{t}\rangle\langle\psi_{t}|-\rangle$, for a typical
trajectory followed by the state vector. The parameter $\gamma$ has
been taken equal to $0.05$ s$^{-1}$, while as usual $\omega = 1$
s$^{-1}$. b) and c) show the average values over 5 and 100
different trajectories, respectively. Note the decoherence effect:
as time increases, the imaginary parts pick up different random
phases and interfere destructively. d) shows the (theoretical)
average values, corresponding to the time evolution of the real
($y$) and imaginary ($z$) parts of the off--diagonal element
$\langle +| \rho(t) |-\rangle$ of the density matrix.} \label{low}
\end{center}
\end{figure}

\begin{figure}
\begin{center}
\makebox[4.2cm]{{\includegraphics[scale=0.23]{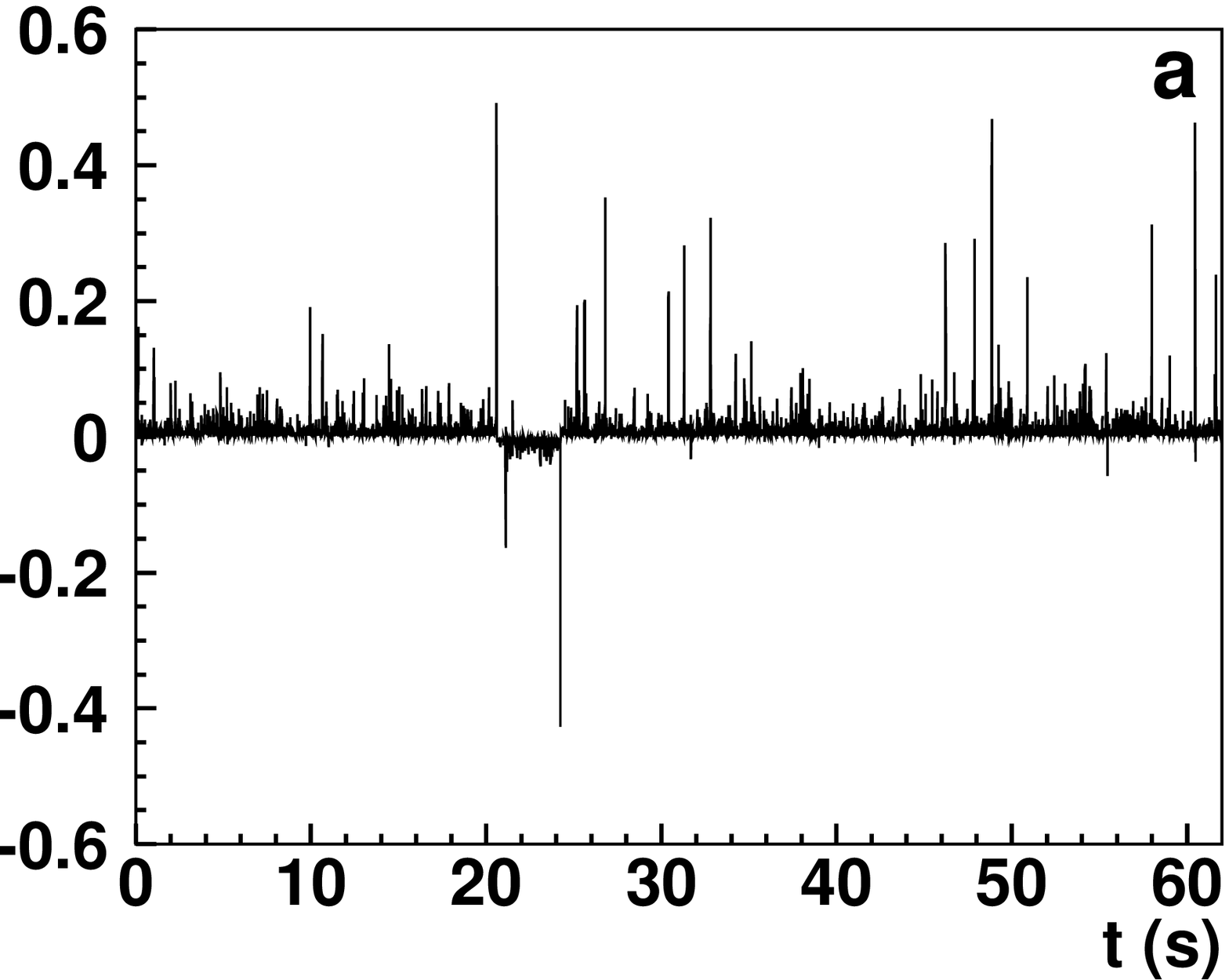}}}
\makebox[4.2cm]{{\includegraphics[scale=0.23]{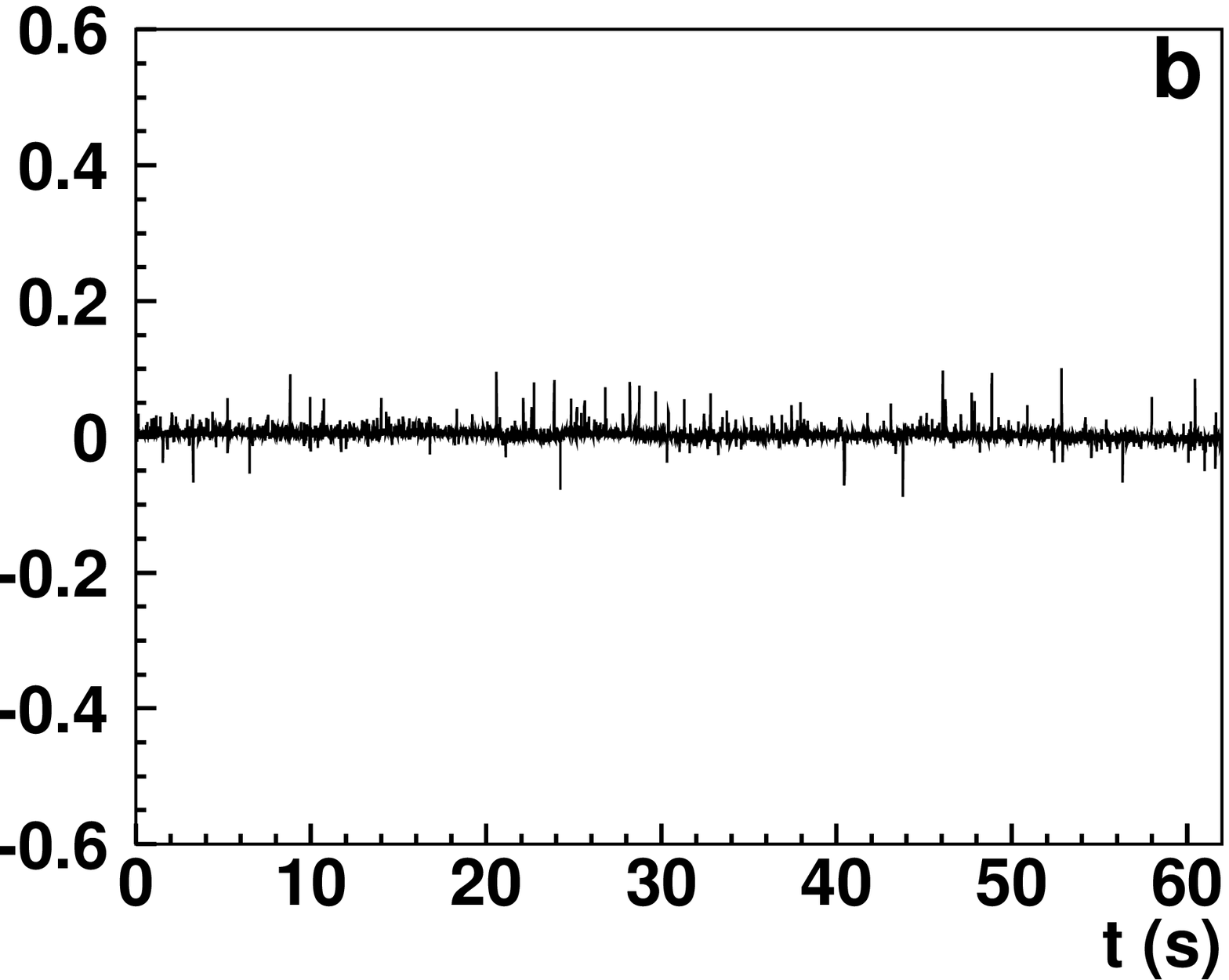}}}
\makebox[4.2cm]{{\includegraphics[scale=0.23]{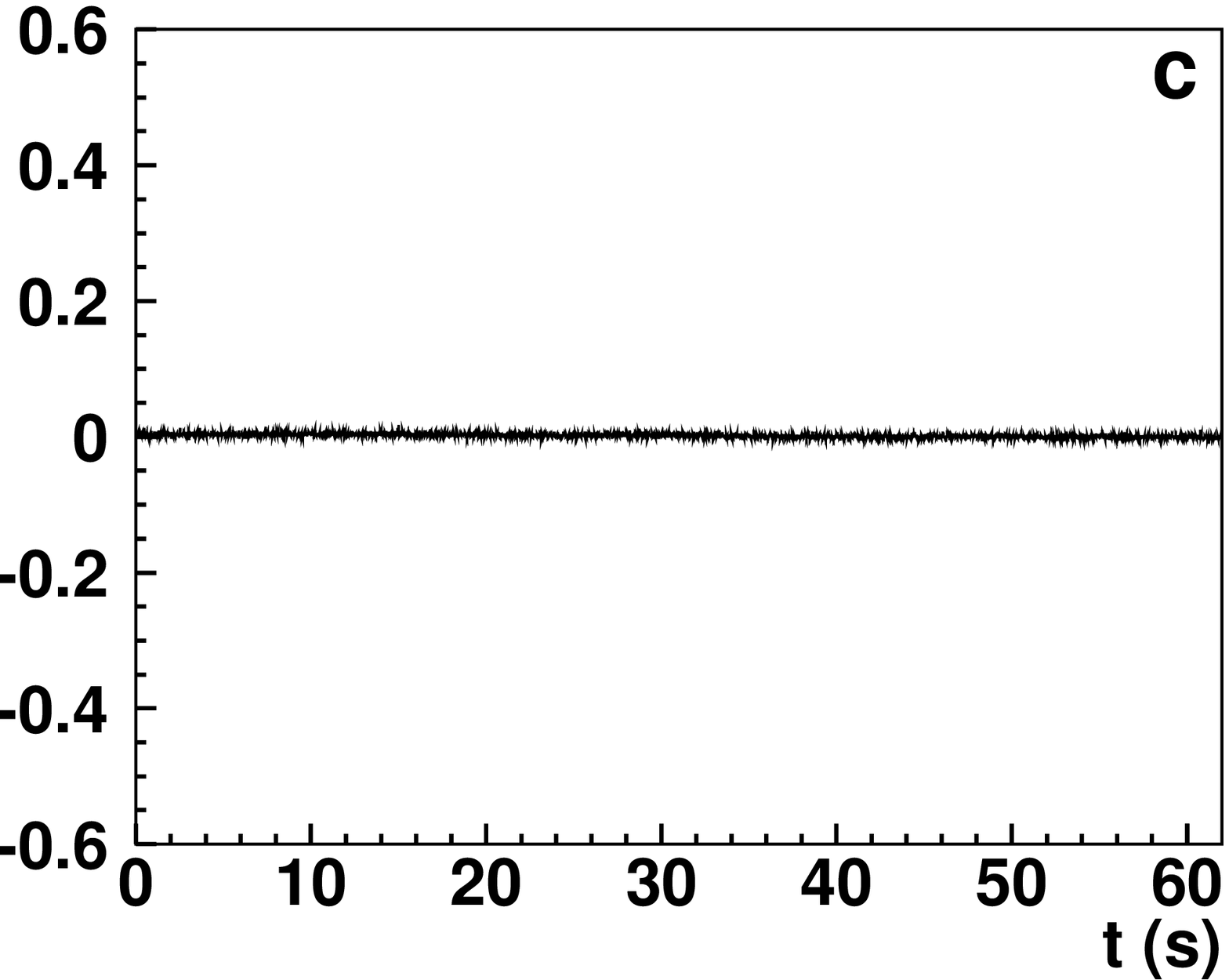}}}
\makebox[4.2cm]{{\includegraphics[scale=0.23]{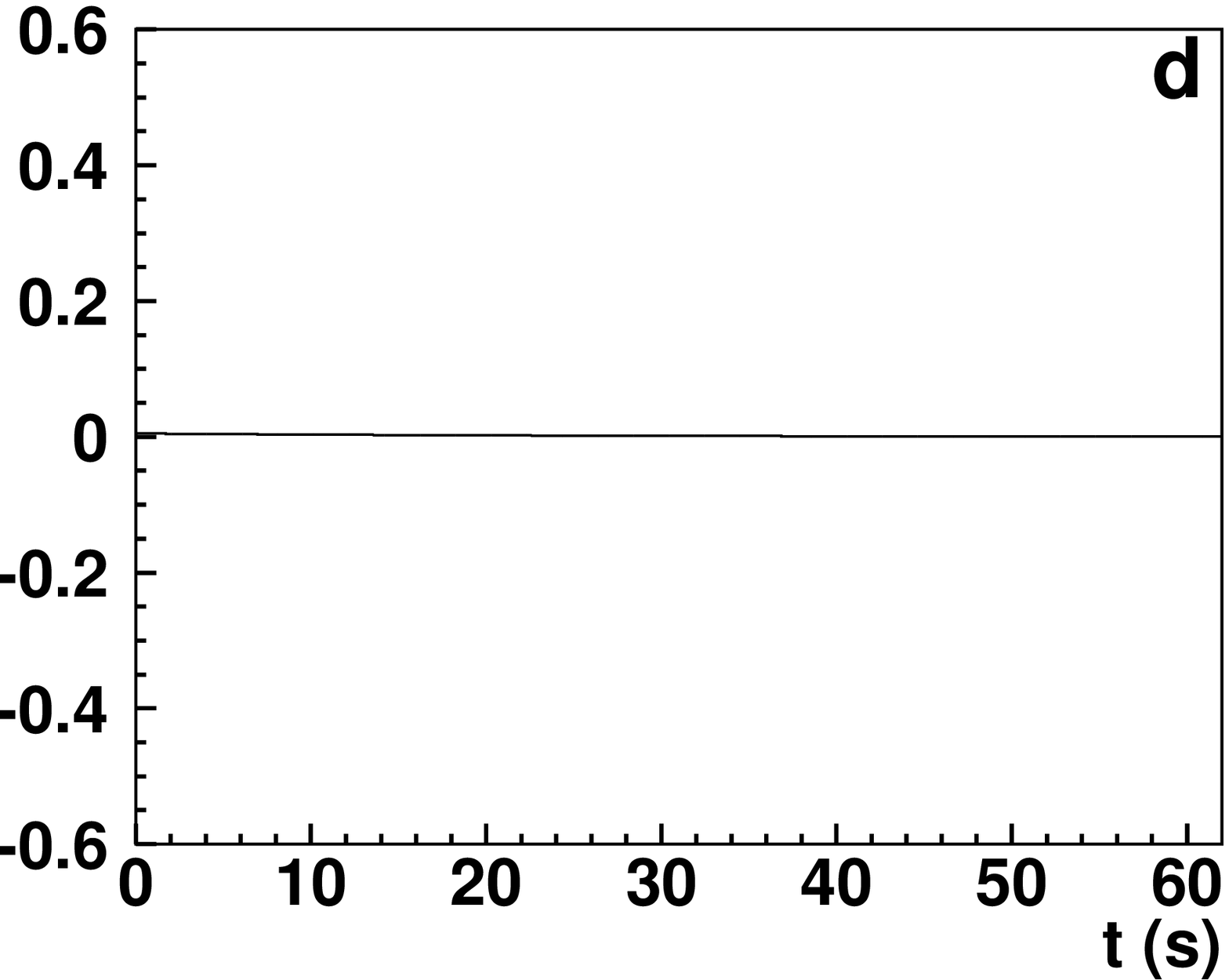}}}
\caption{As in Fig. \ref{low}, but with $\gamma = 50$ s$^{-1}$.
Only the imaginary parts can be seen, as the real ones go rapidly
to zero. Note that the decoherence effect is almost absent: when
$\gamma$ is sufficiently larger than $\omega$, single trajectories
share the property that both the real and the imaginary parts of
$\langle + |\psi_{t}\rangle\langle\psi_{t}|-\rangle$ are (most of
the time) close to zero.} \label{high}
\end{center}
\end{figure}

This property of eq. (\ref{sm}) is well represented in Figs.
\ref{low} and \ref{high}.
Fig. \ref{low}a shows a typical trajectory for the real and
imaginary parts of $\langle +
|\psi_{t}\rangle\langle\psi_{t}|-\rangle$, when $\gamma = 0.05$
s$^{-1}$: the real part (dotted line) decreases in time, while the
imaginary part (continuous line) oscillates and does not approach
the value zero. Figs. \ref{low}b and \ref{low}c show the average
values over 5 and 100 trials, respectively; while the time
evolution of the real part of $\langle +
|\psi_{t}\rangle\langle\psi_{t}|-\rangle$ does not change in a
significant way, the evolution of the imaginary part displays a
typical decoherence effect: the trajectories pick up different
random phases and interfere destructively. Decoherence is then
responsible of the exponential damping of the off--diagonal
element $\langle + |\rho(t)|-\rangle = {\mathbb E}\, [\langle +
|\psi_{t}\rangle\langle\psi_{t}|-\rangle]$ of the statistical
operator, as shown in Fig. \ref{low}d.

Fig. \ref{high}a, instead, shows the result of the numerical
simulations for the (typical) evolution of the real and imaginary
parts of $\langle + |\psi_{t}\rangle\langle\psi_{t}|-\rangle$,
when $\gamma = 50$ s$^{-1}$: both parts are almost always close
zero, as an effect of the reduction process. Figs. \ref{high}b and
\ref{high}c show an average over 5 and 100 trials respectively,
while Fig. \ref{high}d shows the mean value $\langle +
|\rho(t)|-\rangle = {\mathbb E}\, [\langle +
|\psi_{t}\rangle\langle\psi_{t}|-\rangle]$: in this regime, the
damping of the off--diagonal elements of the density matrix is not
due to a decoherence effect but to the reduction process, since
(almost all) realizations of $\langle +
|\psi_{t}\rangle\langle\psi_{t}|-\rangle$ are vanishingly small.

\section{Probabilities}

The second interesting quantity to analyze is the probability that
the state vector reduces toward either the eigenstate $|+\rangle$
or the eigenstate $|-\rangle$. The results of the simulation are
shown in table \ref{tab1}. When $\gamma$ is much larger than
$\omega$, the probability that a reduction occurs approaches the
quantum probabilities of finding the value $+1$ or $-1$ in a
measurement of the observable $\sigma_{z}$, given the initial
state vector (\ref{ivs}):
\begin{eqnarray}
\makebox{Prob.[reduction near $|+\rangle$]} & \simeq & |\langle +
|\psi_{0}\rangle|^2 \; = \; 3/4 \nonumber \\
\makebox{Prob.[reduction near $|-\rangle$]} & \simeq & |\langle +
|\psi_{0}\rangle|^2 \; = \; 1/4 \nonumber
\end{eqnarray}
\begin{table}
\begin{center}
\begin{tabular}{|c|r|r|c|} \hline
$\gamma$ (s$^{-1}$) & red. near $|+\rangle$ & red. near
$|-\rangle$ & total
red. \\ \hline 100 & 73879 (74\%) & 26121 (26\%) & 100000 (100\%) \\
\hline 80 & 74076 (74\%) & 25924 (26\%) & 100000 (100\%) \\ \hline
60 & 73748 (74\%) & 26252 (26\%) & 100000 (100\%) \\ \hline 40 &
73142 (73\%) & 26858 (27\%) & 100000 (100\%) \\ \hline 20 & 70411
(71\%) & 29033 (29\%) & 99444 (99\%) \\ \hline 10 & 45862 (66\%)
& 23279 (34\%) & 69141 (69\%)\\
\hline 5 & 3432 (61\%) & 2165 (39\%) & 5597 (6\%)\\ \hline
\end{tabular}
\caption{Reduction probabilities as a function of $\gamma$. The
last column shows the number of times (over 100000 trials) the
state vector has been reduced within the time interval $[0, 2 \pi]$
seconds. The second and third columns show the numbers of
reductions toward the eigenstate $|+\rangle$ (second column) and
$|-\rangle$ (third column). } \label{tab1}
\end{center}
\end{table}
This is an expected result. In fact, by denoting $\alpha_{t} =
\langle + | \psi_{t} \rangle$ and $\beta_{t} = \langle - |
\psi_{t} \rangle$, one has from eq. (\ref{sm}):
\begin{equation} \label{efa}
d |\alpha_{t}|^2 = -2\omega \makebox{Im}[\alpha_{t}
\beta^{\star}_{t}] dt \, + \, 4 \sqrt{\gamma} |\alpha_{t}|^2 ( 1 -
|\alpha_{t}|^2) dW_{t}
\end{equation}
and similarly for $|\beta_{t}|^2$. In the regime $\gamma \gg
\omega$, one can neglect $\omega$:
\begin{equation} \label{spe}
|\alpha_{t}|^2 = |\alpha_{0}|^2 + 4\sqrt{\gamma} \int_{0}^{t}
|\alpha_{s}|^2 ( 1 - |\alpha_{s}|^2) dW_{s}.
\end{equation}
Since the integrand in the above equation is bounded between 0 and
1, one can conclude \cite{arn} that $|\alpha_{t}|^2$ is a
martingale, which in particular implies that the expectation value
${\mathbb E}\, [\,|\alpha_{t}|^2]$ does not change in time:
\begin{equation}
{\mathbb E}\, [\,|\alpha_{t}|^2] \quad = \quad |\alpha_{0}|^2;
\end{equation}
in other words, reductions (and we have seen that in this regime
all trajectories get reduced) occur with the correct quantum
probabilities.

On the other hand, when $\gamma$ is not much larger than $\omega$,
the approximation $\omega = 0$ is no longer valid, and the
martingale property does not hold any more. As a matter of fact,
the numerical simulation shows that, when $\gamma$ decreases, the
reduction probabilities approach the values
\begin{eqnarray}
\makebox{Prob.[reduction near $|+\rangle$]} & \simeq &  1/2 \nonumber \\
\makebox{Prob.[reduction near $|-\rangle$]} & \simeq &  1/2;
\nonumber
\end{eqnarray}
anyway, we have seen in the previous section that for decreasing
values of $\gamma$, fewer and fewer trajectories are localized, so
that the reduction mechanism progressively loses its statistical
meaning and eventually disappears.

As a final note, we observe that the term $-2\omega
\makebox{Im}[\alpha_{t} \beta^{\star}_{t}] dt$ in Eq. (\ref{efa})
--- which spoils the martingale property of $|\alpha_{t}|^2 $ ---
appears because the standard Hamiltonian $H = \hbar \omega
\sigma_{x}$ and the reduction operator $\sigma_{z}$ do not
commute. If they commuted, such a term would be zero and
$|\alpha_{t}|^2 $ would be a martingale for any value of $\gamma$.

\section{Delocalization time}

Even when $\gamma \gg \omega$, the fact that the Hamiltonian $H$
does not commute with the reduction operator $\sigma_{z}$ has an
important effect, which has not always been stressed in the
literature: once the state vector has reached one eigenstate, it
does not fluctuate around it forever, as there is always the
chance that it jumps to the other eigenstate of $\sigma_{z}$. When
this occur, we speak of a {\it delocalization}.

This behavior of Eq. (\ref{sm}) can be understood by looking at
the equations of motion for the two components $\alpha_t = \langle
+|\psi_t\rangle$ and $\beta_t = \langle -|\psi_t\rangle$:
\begin{eqnarray}
d\alpha_t & = & -i\omega\beta_t dt - 2\gamma\alpha_t (1 -
|\alpha_t|^2)^2dt \nonumber \\
& & + 2 \alpha_t (1 - |\alpha_t|^2)dW_t, \label{eq1} \\
d\beta_t & = & -i\omega\alpha_t dt - 2\gamma\beta_t (1 -
|\beta_t|^2)^2dt \nonumber \\
& & - 2 \beta_t (1 - |\beta_t|^2)dW_t; \label{eq2}
\end{eqnarray}
for simplicity, suppose that the initial state vector
$|\psi_0\rangle$ is equal
--- for example --- to $|+\rangle$ or, equivalently, that the
state vector has been previously reduced to the state $|+\rangle$.

We first of all point out that, if the Hamiltonian $H$ commuted
with the localization operator $\sigma_z$, and in particular if
$H$ were equal\footnote{Which amounts to saying that, in Eq.
(\ref{eq1}), the terms $-i\omega\beta_t$ should be replaced by
$-i\omega\alpha_t$ and, in Eq. (\ref{eq2}), the term
$-i\omega\alpha_t$ should be replaced with $i\omega\beta_t$.} to
$\sigma_z$, there would be no delocalization from the state
$|+\rangle$, as $\alpha_t = e^{-i\omega t}$ and $\beta_t = 0$
would be the stationary solutions of the corresponding equations
of motion. Accordingly, the fact that, in our model, $H$ and
$\sigma_z$ do not commute, is crucial for the delocalization
process.

Coming back to Eqs. (\ref{eq1}) and (\ref{eq2}), an easy way to
explain the delocalization mechanism is to look at numerical
simulations. The noise $dW_t$ has a Gaussian distribution with
zero mean and variance $dt$, so on the average it takes with the
same probability both positive and negative values.

Anyway, it can happen that --- during a certain time interval
--- $dW_t$ takes (almost) only positive (or negative) values: it
is precisely in this conditions that a delocalization may occur,
as it is exemplified in table \ref{tab2}.
\begin{table}
\begin{center}
\begin{tabular}{|c|c|c|} \hline
Step & $|\alpha|^2 = |\langle+|\psi\rangle|^2$ & $dW_t$ \\
\hline \;\;43937\;\; & \;\;0.98595512\;\; & \;\;-0.01479509\;\; \\
\hline 43938 & 0.97223624 & -0.02990961 \\
\hline 43939 & 0.91184014 & -0.04996265 \\
\hline 43940 & 0.87878894 & -0.01746342 \\
\hline 43941 & 0.80806981 & -0.02214550 \\
\hline 43942 & 0.75135818 & -0.01448057 \\
\hline 43943 & 0.55903192 & -0.02797212 \\
\hline 43944 & 0.32168310 & -0.02466131 \\
\hline 43945 & 0.07469396 & -0.03096076 \\
\hline 43946 & 0.06038961 & \; 0.00334229 \\
\hline
\end{tabular}
\caption{Piece of numerical simulation of Eq. (\ref{sm}). The
values of the parameters are: $\omega = 1$ s$^{-1}$, $\gamma =
100$ s$^{-1}$, $dt = 10^{-3}$ s, and $|\psi_0\rangle = |+\rangle$.
The table shows that, when $dW_t$ takes negative values for a
sufficient number of steps, the state vector is driven away from
the state $|+\rangle$ and reduced to the state $|-\rangle$.}
\label{tab2}
\end{center}
\end{table}
There, we have simulated a solution of the equations of motions,
using the following values: $\omega = 1$ s$^{-1}$, $\gamma = 100$
s$^{-1}$, $dt = 10^{-3}$ s, $\alpha_{0} = 1$ and $\beta_{0} = 0$.
The first delocalization begins at the 43937th step: we see from
the table that the noise takes only negative values, thus
determining the collapse of the state vector to the other
eigenstate.

We now move to analyze the delocalization time. It has been
defined in the following way: assume that the state vector has been
reduced to the eigenstate $|+\rangle$, which amounts to saying
that $|\langle + | \psi_{t}\rangle |^2$ remained larger than $1 -
\varepsilon$ for a time interval at least equal to $\tau$. If,
starting from any time $t_{d}$ subsequent to $t_{r}$, the square
modulus $|\langle + | \psi_{t}\rangle |^2$ becomes {\it smaller}
than $1 - \varepsilon$ for at least a time interval equal to
$\tau$, we say that the state vector has been delocalized from the
eigenstate $|+\rangle$ and we call $t_{d}$ the delocalization
time.

\begin{figure}
\begin{center}
{\includegraphics[scale=0.6]{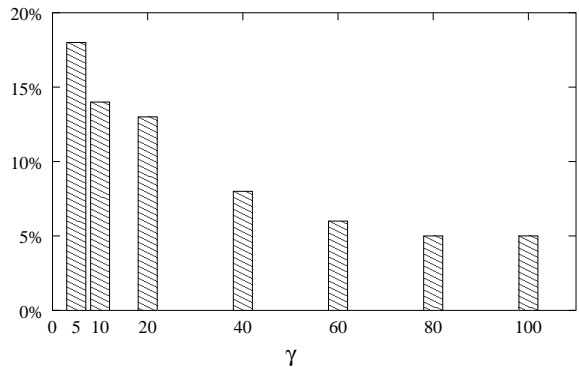}} \caption{The picture
shows  --- as a function of $\gamma = 5,10,20,40,60,80,100$
s$^{-1}$ --- the fraction of all trajectories (which have suffered
a localization) that have been delocalized within the time
interval $[t_{r}, 2\pi]$ seconds.} \label{tdec}
\end{center}
\end{figure}

Fig. \ref{tdec} shows the results of the numerical simulation for
the delocalization times.
We have considered all trajectories which have been localized
within the time interval $[0,2\pi]$ seconds, and we have counted
the number of them which have gone through a delocalization within
the time interval $[t_{r}, 2\pi]$, where $t_{r}$ is the reduction
time for each single trajectory. As expected, for increasing
values of $\gamma$, the fraction of all trajectories which are
delocalized before time $t = 2\pi$ seconds, decreases: once the
state vector gets localized to any of the two eigenstates of
$\sigma_{z}$, it fluctuates around it for a time interval which is
longer, the greater the value of $\gamma$.

\section{Transition from the ``microscopic'' to the
``macroscopic'' regime}

We conclude the analysis of Eq. (\ref{sm}) by giving a general
picture of how the evolution of $|\psi_{t}\rangle$ changes when
moving from the ``microscopic'' ($\gamma \ll \omega$) regime to
the ``macroscopic'' ($\gamma \gg \omega$) one\footnote{In collapse
models \cite{gpr}, the parameter $\gamma$ is constant and much
smaller than the inverse of the characteristic times of the free
Hamiltonian, so that microscopic systems are weakly coupled to the
stochastic terms; this is the reason why the $\gamma \ll \omega$
regime has been called ``microscopic''. On the other hand, it can
be proved \cite{gpr} that macroscopic objects strongly interact
with the noise terms: in fact, the effects of such terms on the
constituents of the macro--system sum with each other, in such a
way that the dynamics of the center of mass of the object is
governed by an effective single--particle collapse--equation, with
a $\gamma$ much greater than the inverse of the characteristic
times of the free Hamiltonian. It is precisely this strong
interaction which is responsible for their classical behavior.
Accordingly, the $\gamma \gg \omega$ regime has been called
``macroscopic''.}.
\begin{figure}
\begin{center}
\makebox[4.2cm]{{\includegraphics[scale=0.5]{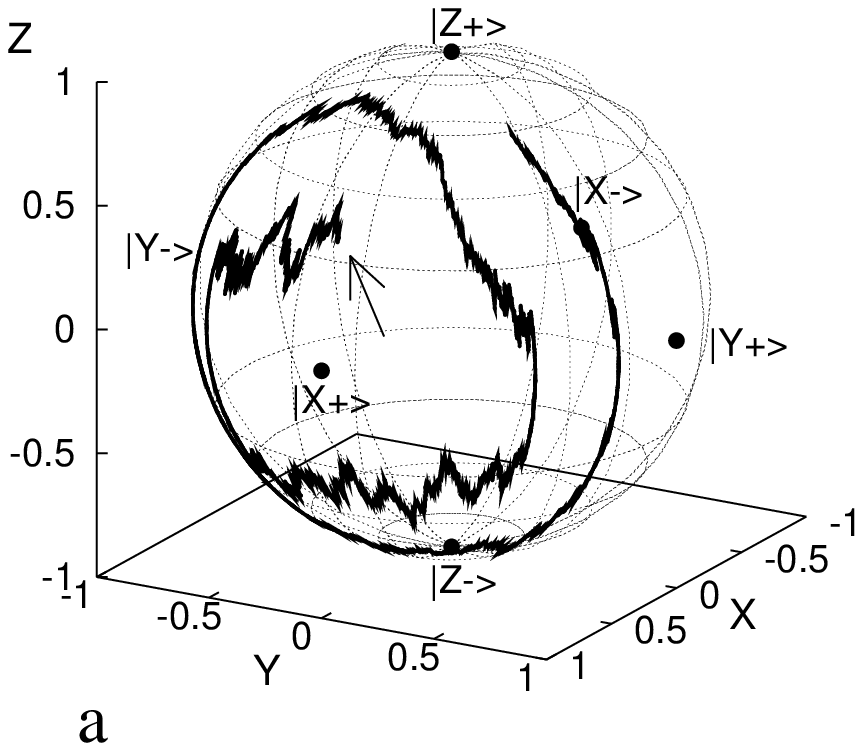}}}
\makebox[4.2cm]{{\includegraphics[scale=0.5]{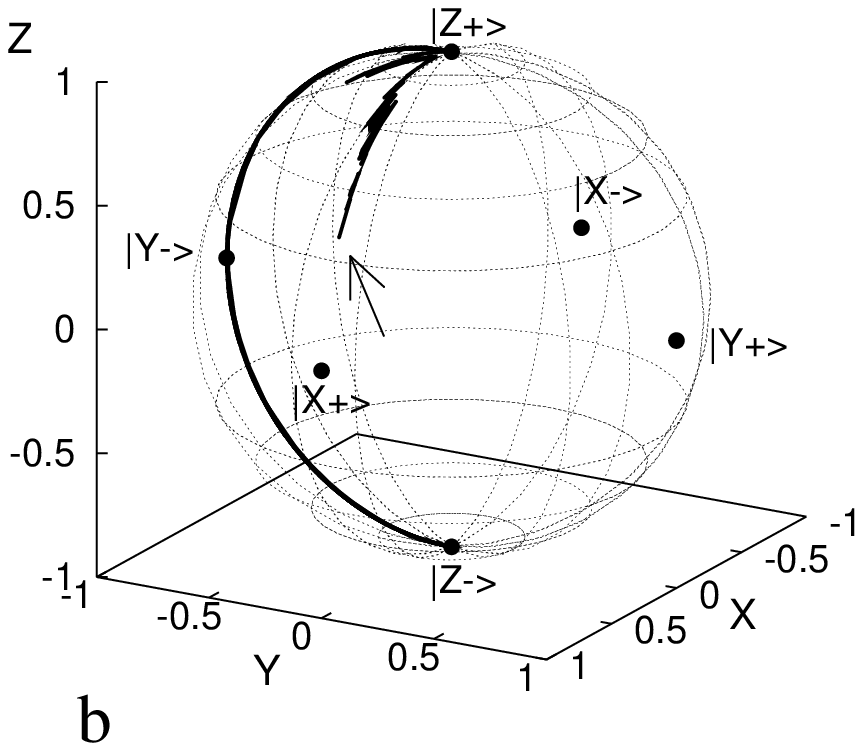}}}
\caption{The figures show two typical trajectories followed by the
state vector in the Bloch sphere, for $\gamma = 0.05$ s$^{-1}$
(Fig. a) and $\gamma = 5$ s$^{-1}$ (Fig. b). In the first case,
the Hamiltonian $H$ rotates the state vector along the $x$ axis of
the Bloch sphere, while the reducing terms slowly move the
rotation plane towards the $|+\rangle$ and $|-\rangle$
eigenstates. In the second case, the effect of the reduction terms
is so strong that the state vector is rapidly driven towards the
$|+\rangle$ eigenstate and then fluctuates around it, until it
jumps to the other eigenstate $|-\rangle$. The arrows mark the
starting point of the trajectories.} \label{ex}
\end{center}
\end{figure}
When $\gamma$ is vanishing small, the state vector simply rotates
along the $x$--axis of the Bloch sphere, guided by the standard
Hamiltonian $H = \hbar \omega \sigma_{x}$. As $\gamma$ increases
--- still remaining smaller than $\omega$ --- $|\psi_{t}\rangle$
keeps rotating along the $x$--axis, but the stochastic terms shift
the rotation plane towards the two eigenstates of $\sigma_{z}$
(see fig. (\ref{ex}a)).

When $\gamma$ becomes significantly greater than $\omega$, the
state vector does not rotate any more around the $x$--axis of the
Bloch sphere: it gets rapidly reduced to one of the two eigenstate
of $\sigma_{z}$ (see fig. (\ref{ex}b)); moreover, if $\gamma$ is
sufficiently large, the reductions occur with the correct quantum
probabilities.

The state vector then fluctuates around the eigenstate to which it
has been reduced for a time interval that --- statistically --- is
longer, the greater the values of $\gamma$; accordingly, when
$\omega$ can be neglected, the system behaves like a macroscopic
system, regarding the physical properties of the spin along the
$z$-axis: they always have a well--definite value.

\section{Comparison with space--collapse models}

Collapse models have been proposed in order to solve the
measurement problem of quantum mechanics, and in particular to
provide an explanation to the fact that macroscopic objects always
have a definite position in space: accordingly, in most proposals
on spontaneous collapse the localization operators are functions
of the position operators\footnote{In refs \cite{ad1,ad2}, an
energy--based collapse model has been formulated; in this case,
the Hamiltonian commutes with the reduction operators, so the
model is different from the one considered here.}. In this
section, we confront the dynamics described by Eq. (\ref{sm}) with
the properties of the collapse model proposed in ref. \cite{gpr}.
The stochastic Schr\"odinger equation of this model takes the
form:
\begin{eqnarray} \label{fse}
d|\psi_{t}\rangle & = & \left[ -\frac{i}{\hbar}H dt  +
\sqrt{\gamma} \int d^{3}x \left( N({\bf x}) - \langle N_{t}({\bf
x}) \rangle \right)dW_{t}({\bf x}) \right. \nonumber \\
& & \left. - \frac{\gamma}{2} \int d^{3}x \left( N({\bf x}) -
\langle N_{t}({\bf x}) \rangle \right)^{2} dt \, \right]
|\psi_{t}\rangle,
\end{eqnarray}
where $\langle N_{t}({\bf x}) \rangle  = \langle \psi_{t}| N({\bf
x}) |\psi_{t}\rangle$ and $H$ is the standard Hamiltonian for the
system under study. The ``average density number operator''
$N({\bf x})$ is defined as follows:
\begin{equation}
N({\bf x}) = \left( \frac{\alpha}{2\pi}\right)^{3/2} \sum_{s} \int
d^{3}y\, e^{\displaystyle -\frac{\alpha}{2}({\bf x} - {\bf
y})^{2}} a^{\dagger}({\bf y}, s)\, a({\bf y}, s),
\end{equation}
$a^{\dagger}({\bf y}, s)$ and $a({\bf y}, s)$ being, respectively,
the creation and annihilation operators of a particle at point
${\bf y}$ with spin $s$; $W_{t}({\bf x})$ is a family of
independent Wiener processes: ${\mathbb E}[ dW_{t}({\bf x})] = 0$,
${\mathbb E}[ dW_{t}({\bf x})\, dW_{t}({\bf y})] =
\delta^{(3)}({\bf x} - {\bf y})\, dt$.

Finally, the two parameters, $\alpha$ and $\gamma$, appearing in
the model, have been taken equal to:
\begin{equation} \label{vp}
\frac{1}{\sqrt{\alpha}} \, = \, 10^{-5}\, \makebox{cm}, \quad
\gamma \, = \, \lambda \left( \frac{4\pi}{\alpha}
\right)^{\frac{3}{2}}, \quad \lambda \, = \, 10^{-17}\;
\makebox{s$^{-1}$}.
\end{equation}

Eq. (\ref{fse}) has the same structure of Eq. (\ref{sm}) and, in
all interesting physical situations, the Hamiltonian $H$ does not
commute with the localization operators $a^{\dagger}({\bf y}, s)\,
a({\bf y}, s)$: as a consequence the dynamics described by Eq.
(\ref{fse}) is, qualitatively, analogous to that described by Eq.
(\ref{sm}). We comment on the implications of our analysis of the
spin--model (\ref{sm}), for the space--localization model given by
Eq. (\ref{fse}).

\noindent {\bf Reduction mechanism.} In the case of a free object,
the analog of the time $\omega^{-1}$ can be reasonably identified
with the time $T$ it takes for the spread $\Delta x$ of the center
of mass of the object to change appreciably, e.g. to double its
initial value; in the case of a free particle of mass $m$ (in one
dimension), we have:
\begin{equation}
\Delta x(t) \; = \; \Delta x(t_{0})\, \sqrt{1 + \frac{h^2 [\Delta
x(t_{0})]^{-4}}{4 m^2}\, t^2},
\end{equation}
so that $\Delta x(t) = 2 \Delta x(t_{0})$ when $t = T$, with:
\begin{equation}
T = \frac{\sqrt{12} m [\Delta x (t_{0})]^2}{h}.
\end{equation}
In the above formula, we shall take $\Delta x (t_{0}) = 10^{-5}$
cm, which is the typical localization length of the model (compare
with eq. (\ref{vp})).

According to the numerical values given to the parameters $\alpha$
and $\gamma$, in the microscopic situation (when only one or a few
constituents are present) we are in the situation where the
characteristic time $T$ associated to the Hamiltonian is
exceedingly smaller than the characteristic time $\lambda^{-1}$
associated to the reduction mechanism; for a particle like a
proton, we have, e.g.:
\begin{equation}
T \; \sim \; 10^{-7}\, \makebox{s} \quad \ll \quad \lambda^{-1} \;
\simeq \; 10^{17}\, \makebox{s}.
\end{equation}
In this situation, the standard quantum dynamics of microscopic
systems in not appreciably altered by the presence of the
stochastic terms.

On the other hand, in the macroscopic case, as already mentioned,
an ``amplification mechanism'' occurs: it can be shown \cite{gpr}
that the frequency $\Lambda$ associated to the reduction of the
center of mass of a rigid body is approximately given by the
formula: $\Lambda = N \lambda$, where $N$ is the number of
constituents of the object. For an object of mass equal to 1 gr
($N \sim N_{\makebox{\tiny Avogadro}}$), we have:
\begin{equation}
T \; \sim \; 10^{17}\, \makebox{s} \quad \gg \quad \Lambda^{-1} \;
\sim \; 10^{-7}\, \makebox{s}.
\end{equation}
Accordingly, the characteristic time $T$ associated to a
macroscopic object is many order of magnitudes greater than
$\Lambda^{-1}$. In this situation, it can be proved
that\footnote{See ref. \cite{rep} and references therein for all
calculations and details.}:

\noindent a) The reduction time is extremely well defined and
small ($\sim 10^{-6}$ s), much smaller than the perception time of
a human being. All macroscopic objects are localized in a very
fast way, and superpositions of different macroscopic states
cannot persist.

\noindent b) After the reduction as occurred, the wavefunction of
the system is very well localized in space: for typical
macroscopic objects its spread in position is $\sim 10^{-11}$ cm.

Of course, the wavefunction cannot have a compact support, since
the Hamiltonian (when it does not commute with the position
operators) would immediately turn it into a wavefunction with
``tails'' going off to infinity; anyway, the fraction of total
wavefunction corresponding to the tails is of the order of
$e^{-10^{18}}$, which means that they can be neglected for all
practical purposes.

\noindent {\bf Probabilities.} The reduction probabilities are
almost exactly equal to the quantum probabilities and the outcomes
of measurement processes follow the Born rule. Of course, the
physical predictions of Eq. (\ref{fse}) are in principle different
from those of standard quantum mechanics: this opens the way to
the possibility of testing collapse models, especially at the
mesoscopic level where the time scales of the quantum dynamics and
the time scales of the reduction mechanism become comparable.
Anyway, such kind of tests are difficult to implement as it is
difficult to control all sources of noise which mask the effects
of spontaneous localizations.

\noindent {\bf Delocalization process.} We have seen in section IV
that, when the spin--wavefunction has been localized into one of
the two eigenstates of $\sigma_{z}$, it will sooner or later be
driven by the dynamics of Eq. (\ref{sm}) into the other eigenstate
and back. The dynamics of Eq. (\ref{fse}) displays a similar
behavior.

After a macroscopic object has been localized in space (take its
velocity equal to zero), there is a non zero probability that it
gets delocalized and, shortly after that, localized again in a
different region of space. Anyway, the probability that such an
event occurs is exceedingly small\footnote{And, in particular,
many orders of magnitude longer than the age of the universe.}
(practically equal to zero), being proportional to the fraction of
the wavefunction corresponding to its tails. Accordingly, when a
macroscopic object gets localized, it stays localized practically
forever.

Note also that this feature of space collapse is common to
standard quantum mechanics as well: any macroscopic object, being
described by a wavefunction with noncompact support, has a
(vanishing small) probability of going through a quantum jump from
a space region to a far away one.

\section*{Acknowledgements}

We acknowledge many useful discussions with Prof. G.C. Ghirardi.
We also thank Prof. S. Baroni for having introduced us to the
field of numerical simulations of stochastic differential
equations.

\section*{Appendix I: Complete solutions of the equation for the
statistical operator}

The solutions of equations (\ref{eqxyz}) are given as follows:

\noindent {\bf Over--damped case: $\gamma > 2\omega$}.
\begin{eqnarray}
x(t) & = & a\,e^{-\gamma(1+\mu)t} \; + \; b\, e^{-\gamma(1-\mu)t}
\; + \; 1/2 \nonumber \\
y(t) & = & y_{0}\,e^{-2\gamma t} \nonumber \\
z(t) & = & c\,e^{-\gamma(1-\mu)t} \; + \; d\, e^{-\gamma(1+\mu)t},
\end{eqnarray}
with:
\begin{eqnarray}
a & = & -\frac{2\omega^2}{\gamma^{2}\mu(1+\mu)}\, x_{0} \; + \;
\frac{\omega}{\gamma\mu}\, z_{0} \; + \;
\frac{\omega^2}{\gamma^{2}\mu(1+\mu)} \nonumber \\
b & = & \frac{2\omega^2}{\gamma^{2}\mu(1-\mu)}\, x_{0} \; - \;
\frac{\omega}{\gamma\mu}\, z_{0} \; - \;
\frac{\omega^2}{\gamma^{2}\mu(1-\mu)} \nonumber \\
c & = & -\frac{\omega}{\gamma\mu}\, x_{0} \; + \;
\frac{1+\mu}{2\mu}\, z_{0} \; + \; \frac{\omega}{2\gamma\mu}
\nonumber \\
d & = & \frac{\omega}{\gamma\mu}\, x_{0} \; - \;
\frac{1-\mu}{2\mu}\, z_{0} \; - \; \frac{\omega}{2\gamma\mu}
\nonumber \\
\mu & = & \sqrt{1- \left(\frac{2\omega}{\gamma}\right)^2}
\end{eqnarray}

\noindent {\bf Critically damped case: $\gamma = 2\omega$}.
\begin{eqnarray}
x(t) & = & [a \, + \, b\,t]\, e^{\displaystyle-\gamma\, t}
\; + \; 1/2 \nonumber \\
y(t) & = & y_{0}\,e^{\displaystyle-2\gamma t} \nonumber \\
z(t) & = & [c \, + \, d\, t ]\,e^{\displaystyle-\gamma\,t},
\end{eqnarray}
with:
\begin{eqnarray}
a & = & x_{0} - \frac{1}{2} \nonumber \\
b & = & (2x_{0} - 1)\omega \, - \, \gamma z_{0} \nonumber \\
c & = & z_{0} \nonumber \\
d & = & (2x_{0} - 1)\omega \, - \, \gamma z_{0}.
\end{eqnarray}

\noindent {\bf Under--damped case: $\gamma < 2\omega$}.
\begin{eqnarray}
x(t) & = & \left[ a \cos (\lambda t) \, + \, b \sin (\lambda t)
\right]e^{\displaystyle-\gamma t} \; + \; 1/2 \nonumber \\
y(t) & = & y_{0}\,e^{\displaystyle-2\gamma t} \nonumber \\
z(t) & = & \left[ c \cos (\lambda t) \, + \, d \sin (\lambda t)
\right]e^{\displaystyle-\gamma t},
\end{eqnarray}
with:
\begin{eqnarray}
a & = & x_{0} - \frac{1}{2} \nonumber \\
b & = & \frac{\gamma}{\lambda}\, x_{0} \, - \,
\frac{2\omega}{\lambda}\, z_{0} -
\frac{\gamma}{2\lambda} \nonumber \\
c & = & z_{0}
\nonumber \\
d & = & \frac{1}{\lambda} \left( 2\omega x_{0} \, - \, \omega \, -
\, \gamma\, z_{0}
\right) \nonumber \\
\lambda & = & \gamma\,\sqrt{\left(\frac{2\omega}{\gamma}\right)^2
-1}
\end{eqnarray}

\section*{Appendix II: A note on the numerical simulation}

\begin{figure}
\begin{center}
{\includegraphics[scale=0.41]{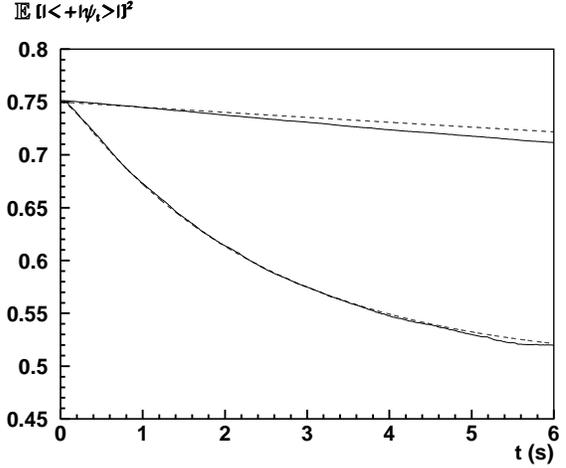}} \caption{The two dotted
curves represent the (analytical) evolution
$\langle+|\rho(t)|+\rangle$, with $\gamma = 100$ s$^{-1}$ (upper
curve) and $\gamma = 5$ s$^{-1}$ (lower curve). The continuous
curves show the average value of $|\langle+|\psi_{t}\rangle|^2$
over 100000 trajectories of $|\psi_{t}\rangle$, for the same
values of $\gamma$. For the values of the time step, see the
text.} \label{eul}
\end{center}
\end{figure}

To solve eq. (\ref{sm}), we have used the simple Euler algorithm,
which proved to be satisfactory for the purposes of the article.
To test the goodness of the simulation, we have
compared\footnote{Of course, such a comparison tests the
convergence of the numerical algorithm in the {\it weak} sense
(i.e. in the mean) \cite{pla}, while a test for the convergence in
the {\it strong} sense (for single trajectories) would have been
better; but this kind of a test is not feasible. Note that Euler
scheme is of order $1$ for the convergence in the weak sense,
while it is of order $1/2$ for the convergence in the strong
sense.} the analytical evolution of ${\mathbb E}\,
[|\langle+|\psi_{t}\rangle|^2] = \langle+|\rho(t)|+\rangle$ with
the average value of $|\langle+|\psi_{t}\rangle|^2$ over many
sample trajectories, obtained by solving eq. (\ref{sm})
numerically (see Fig. \ref{eul}).

While analyzing the data, we have seen that for large values of
$\gamma$ --- i.e. when the reduction mechanism becomes extremely
rapid --- the numerical simulation was not as good as expected.
For such a reason we have decided to decrease the time step during
the first $0.1$ s, obtaining a sensible improvement.

We have made different tests on the choice of the time steps and
number of trials and, taking also into account the time required
by our CPU for elaborating the data, we have found that the best
choice (according to the previously mentioned criterion) was to
take:
\begin{center}
\begin{tabular}{lr}
Time step for $t \leq 0.1$ s: & $10^{-7}$ \\
Time step for $t > 0.1$ s: & $10^{-3}$ \\
Number of trials: & 100000.
\end{tabular}
\end{center}

\end{document}